\documentclass[11pt]{iopart}

\usepackage{epsfig,graphicx,latexsym,iopams}
\usepackage{color}
\usepackage{graphicx}  

\def\msun{\,{\rm M_\odot}}

\newcommand\be{\begin{equation}}
\newcommand\ee{\end{equation}}
\newcommand{\ba}{\begin{eqnarray}}
\newcommand{\ea}{\end{eqnarray}}

\begin{document}

\title[]{Insights on the astrophysics of supermassive black hole binaries from pulsar timing observations}

\author{A Sesana$^1$}

\address{$^1$\ Max-Planck-Institut f\"ur Gravitationsphysik, Albert Einstein Institut, Am M\"ulenber 1, 14476 Golm, Germany}

\begin{abstract}
Pulsar timing arrays (PTAs) are designed to detect the predicted gravitational wave (GW) background produced by a cosmological population of supermassive black hole (SMBH) binaries. In this contribution I review the physics of such GW background, highlighting its dependence on the overall binary population, the relation between SMBHs and their hosts, and their coupling with the stellar and gaseous environment. The latter is particularly relevant when it drives the binaries to extreme eccentricities ($e>0.9$), which might be the case for stellar-driven systems. This causes a substantial suppression of the low frequency signal, potentially posing a serious threat to the effectiveness of PTA observations. A future PTA detection will allow to directly observe for the first time subparsec SMBH binaries on their way to the GW driven coalescence, providing important answers of the outstanding questions related to the physics underlying the formation and evolution of these spectacular sources.  
\end{abstract}

\section{introduction}
The pulsar timing arrays (PTAs) described in this volume, provide a unique opportunity to obtain the very first low-frequency gravitational wave (GW) detection. The European Pulsar Timing Array (EPTA) \cite{ferdman10}, the North American Nanohertz Observatory for Gravitational Waves (NANOGrav) \cite {jenet2009}, and the Parkes Pulsar Timing Array (PPTA) \cite{man12}, joining together in the International Pulsar Timing Array (IPTA) \cite{hobbs2010}, are constantly improving their sensitivity in the frequency range of $\sim10^{-9}-10^{-6}$ Hz.  Inspiralling supermassive black hole (SMBH) binaries populating merging galaxies throughout the Universe are expected to generate the dominant signal in this frequency band \cite{rr95,jaffe03,wl03,sesana08}. 

Despite the fact that theoretical models of galaxy formation in the standard hierarchical framework predict a large population of SMBH binaries forming during galaxy mergers, to date there is only circumstantial observational evidence of their existence. Less than 20 SMBH pairs with separations of $\sim10$ pc to $\sim10$ kpc are known to date (see \cite{dotti12}, for a comprehensive review). At smaller separation, only a handful of candidate sub-parsec bound Keplerian SMBH binaries have been identified, based on peculiar broad emission line shifts \cite{tsalmantza11,eracleous11}; however, alternative explanations to the binary hypothesis exist \cite{dotti12}, and unquestionable observational evidence is still missing.

If as abundant as predicted, SMBH binaries are expected to form a low frequency background of gravitational waves (GWs) with a typical strain amplitude $A\sim10^{-15}$ at a frequency $f=1/$yr \cite{wl03,sesana08,ravi12,mcwilliams12}, with a considerable uncertainty of $\approx$0.5dex{\footnote{In astronomy, the notation dex is commonly used for the log$_{10}$ unit; therefore 0.5dex$=10^{0.5}$.}}. The aforementioned studies indicate that such a signal is expected to be dominated by a handful of sources, some of which might be individually resolvable. On one hand, the unresolved background provides innovative ways to test fundamental physics and alternative theories of gravity; on the other hand, electromagnetic counterparts to individually resolvable sources can be searched for with a number of facilities opening new avenues toward a multimessenger based understanding of these fascinating systems and their hosts. These themes are not included in this paper, but are covered in K.J. Lee and T. Tanaka \& Z. Haiman contributions to the present special issue. Here I provide a general overview of the predicted GW signal as a whole, discussing uncertainties in normalization and spectral shape stemming from the underlying properties of the emitting binaries. I will be generally concerned with the {\it level of the background}, without entering into its peculiar properties in terms of non-Gaussianity and resolvability \cite{sesana09,ravi12}, nor in issues related to detection, which are treated in the contributions by X. Siemens and collaborators, J. Ellis and N. Cornish \& A. Sesana. 

The paper is organized as follows. In Section 2, the concept of GW background is introduced, and the relevant ingredients that enter its computation are identified. The main focus of Section 3 is on the overall cosmological population of SMBH binaries (namely their number and typical masses) and on the information that can be extracted by a putative PTA observation. Section 4 is devoted to the coupled dynamical evolution of SMBH binaries and their star/gas rich environment. This coupling has important consequences on the source frequency distribution and their eccentricity, which leaves important signatures in the signal. A summary of the main results is given in Section 5. Throughout the paper a concordance $\Lambda$--CDM universe with $\Omega_M=0.27$, $\Omega_\lambda=0.73$ and $h=0.7$ is assumed. Unless otherwise specified, equations are casted in geometric units where $G=c=1$.

\section{General model of the GW background}
Consider a cosmological population of merging SMBH binaries. Each merging pair is characterized by the masses of the two holes $M_1>M_2$, defining the mass ratio $q=M_2/M_1$. Without making any restrictive assumption about the physical mechanism driving the binary semimajor axis and eccentricity evolution, we can write the characteristic amplitude $h_c$ of the GW signal generated by such population as:
\begin{eqnarray}
h_c^2(f) = &\int_0^{\infty}dz\int_0^{\infty}dM_1\int_0^{1}dq \frac{d^4N}{dzdM_1dqdt_r}\frac{dt_r}{d{\rm ln}f_{{\rm K},r}}\times\nonumber\\
& h^2(f_{{\rm K},r})\sum_{n=1}^{\infty}\frac{g[n,e(f_{{\rm K},r})]}{(n/2)^2}\,\delta\left[f-\frac{nf_{{\rm K},r}}{1+z}\right].
\label{hch2}
\end{eqnarray}
Here, $h(f_{{\rm K},r})$ is the strain emitted by a circular binary at a Keplerian rest frame frequency $f_{{\rm K},r}$, averaged over source orientations
\begin{equation} 
h(f_{{\rm K},r})=\sqrt{\frac{32}{5}}\frac{{\cal M}^{5/3}}{D}(2\pi f_{{\rm K},r})^{2/3},
\label{hrms}
\end{equation}
where we have introduced the chirp mass ${\cal M}=(M_1M_2)^{3/5}/(M_1+M_2)^{1/5}$, and the comoving distance to the source $D$.
The function $g(n,e)$ \cite{pm63} accounts for the fact that the binary radiates GWs in the whole spectrum of harmonics $f_{r,n}=nf_{{\rm K},r}\,\,\,(n=1, 2, ...)$, and is given by, e.g., equations (5)-(7) in \cite{amaro10}. The $\delta$ function, ensures that each harmonic $n$ contributes to the signal at an observed frequency $f=nf_{{\rm K},r}/(1+z)$, where the factor $1+z$ is given by the cosmological redshift. $d^4N/(dzdM_1dqdt_r)$ is the differential cosmological coalescence rate (number of coalescences per year) of SMBH binaries per unit redshift $z$, primary mass $M_1$, and mass ratio $q$, and $dt_r/d{\rm ln}f_{{\rm K},r}$ is the time spent by the binary at each logarithmic frequency interval. These two latter terms, taken together, simply give the instantaneous population of comoving systems orbiting at a given logarithmic Keplerian frequency interval per unit redshift, mass and mass ratio. In the case of circular GW driven binaries, $g(n,e)=\delta_{n2}$, $dt/d{\rm ln}f$ is given by the standard quadrupole formula, and equation (\ref{hch2}) reduces to the usual form  
\begin{equation}
h_c^2(f) =\frac{4f^{-4/3}}{3\pi^{1/3}}\int \int dzd{\cal M} \, \frac{d^2n}{dzd{\cal M}}{1\over{(1+z)^{1/3}}}{\cal M}^{5/3},
\label{hcirc}
\end{equation}
where we have introduced the differential merger remnant density (i.e. number of mergers remnants per co moving volume) $d^2n/(dzd{\cal M})$ (see \cite{phinney01,sesana13} for details). In this case, $h_c\propto f^{-2/3}$; it is therefore customary to write the characteristic amplitude in the form $h_c=A(f/{\rm yr}^{-1})^{-2/3}$, where $A$ is the amplitude of the signal at the reference frequency $f=1 {\rm yr}^{-1}$. Observational limits on the GW background are usually given in terms of $A$.

Equation (\ref{hch2}), together with a prescription for the eccentricity distribution of the emitting SMBH binaries as a function of the frequency, namely $e(M_1,q,f_{{\rm K,r}})$, provides the most general description of the GW background generated by a population of SMBH binaries. The signal depends on three distinctive terms:
\begin{enumerate}
\item the cosmological coalescence rate of SMBH binaries in the Universe, $d^4N/(dzdM_1dqdt_r)$;
\item the specific frequency evolution of each binary, $dt_r/d{\rm ln}f_{{\rm K},r}$;
\item the eccentricity evolution of the systems, which determines the emitted spectrum for a given binary Keplerian frequency.
\end{enumerate}
In the following section, we will examine the impact of the items listed above on the GW signal; on the other hand we will highlight the enormous potential of PTA observations in improving our understanding of the global population of SMBH binaries in our Universe and of their dynamical evolution.

\section{Spectral normalization: cosmological SMBH binary coalescence rate}
As written in equation (\ref{hch2}), the GW strain amplitude is proportional to the square root of the cosmic coalescence rate of SMBH binaries, and it is sensitive to the mass distribution of those binaries. The SMBH binary coalescence rate therefore sets the {\it normalization} of the detectable signal. This, in practice, depends on four ingredients: (i) the galaxy merger rate; (ii) the relation between SMBHs and their hosts, (iii) the efficiency of SMBH coalescence following galaxy mergers and (iv) when and how accretion is triggered during a merger event.

{\it Galaxy merger rate.} Despite the number of observations of massive galaxies at relatively low redshift, their merger rate is not very well constrained, and is one of the major factors of uncertainties in the calculation of the signal. As detailed in \cite{sesana13}, one possible observationally based way to estimate the galaxy differential merger rate is the following:
\begin{equation}
\frac{d^3n_G}{dzdMdq}=\frac{\phi(M,z)}{M\ln{10}}\frac{{\cal F}(z,M,q)}{\tau(z,M,q)}\frac{dt_r}{dz}.
\label{galmrate}
\end{equation}
Here, $\phi(M,z)=(dn/d{\rm log}M)_z$ is the galaxy mass function measured at redshift $z$; ${\cal F}(M,q,z)=(df/dq)_{M,z}$ is the differential fraction of galaxies with mass $M$ at redshift $z$ paired with a secondary galaxy having a mass ratio in the range $[q, q+\delta{q}]$, and $\tau(z,M,q)$ is the typical merger timescale for a galaxy pair with a given $M$ and $q$ at a given $z$. $\phi$ and ${\cal F}$ can be directly measured from observations, whereas $\tau$ can be inferred by detailed numerical simulations of galaxy mergers. All these quantities are known at best to within a factor of 2 , implying that the galaxy merger rate can be estimated within an accuracy of a factor of a few. Alternatively, the galaxy merger rate can be estimated from large N-body simulations of structure formation. Here the problem is that only few of those are available to date (see, e.g., the Millennium run \cite{springel05}), and it is therefore difficult to extract sensible errorbars on the numbers.

{\it SMBH-host relations.} The more massive the SMBH, the stronger the emitted GW signal. Observationally, SMBHs correlate both with the velocity dispersion, $\sigma$, and the mass, $M_{\rm bulge}$, of the host galaxy bulge \cite{mago98,ferrarese00}. However, those correlations come in different flavors, and are constantly re-calibrated to include new available data (see, e.g., \cite{haring04,gultekin09,sani11,mcconnell13,grahamscott13}). Most noticeably, the discovery of SMBHs with $M>10^{10}\msun$ in two brightest cluster galaxies (BCGs) \cite{fabian12} resulted in a recent upward revision to the established SMBH-host relations by a factor 0.2-0.3dex \cite{mcconnell13,grahamscott13}. Moreover, these relations have a significant intrinsic scatter ($\approx$0.3-0.4dex), making a proper determination of the global SMBH mass function and mass density problematic. 

{\it Efficiency of SMBH coalescence.} Even if galaxies merge, the two SMBHs have to make their way to the center of the merger remnant, form a Keplerian binary, and get rid of their energy and angular momentum to enter the efficient GW emission stage. In most of the models underlying current analysis efforts \cite{sesana08,mcwilliams12,ravi12}, the coalescence efficiency is taken into account through the estimate of the Chandrasekhar dynamical friction timescale. If this timescale is longer than the Hubble time, the secondary SMBH never makes it to the center of the merger remnant. Otherwise, a Keplerian binary forms, and its subsequent evolution is assumed to occur quickly on cosmological timescales (e.g., $\lesssim100$Myr). However, the physical problem of the SMBH binary decay both in stellar and gaseous environments has not been completely solved (see \cite{dotti12} for a review). Full N-body simulations indicate that efficient merger occurs in star dominated environments \cite{khan11,preto11}, and the presence of massive circumbinary discs might facilitate the process \cite{escala05,dotti06}. However, simplifying assumptions in the initial conditions and in the treatment of accretion disc physics make the extrapolation from those results to realistic systems uncertain. Nevertheless, we stick here to the standard prescription of quick coalescence after binary formation (for the practical purpose of computing the signal, the coalescence of a binary forming following the merger of a galaxy pair at redshift $z$ occurs at $z-\Delta{z}$, where $\Delta{z}$ is given by the the dynamical friction timescale).

{\it Accretion.} A further factor of uncertainty is related to accretion onto the SMBHs. During galaxy mergers, large amounts of gas are subject to dynamical instabilities and are prone to fall towards the minimum of the evolving potential well \cite{mihos96}, eventually triggering accretion that increases the mass of the SMBHs. Whether this occurs before, during or after the Keplerian binary stage has an effect onto the effective mass and mass ratio of the GW emitting systems. Quantitative estimations by \cite{sesana09} showed that this can have a factor of $\approx2$ impact on the overall GW signal.

\subsection{What can we learn from PTA observations}
\begin{figure}
   \centering
   \includegraphics[width=5.0in]{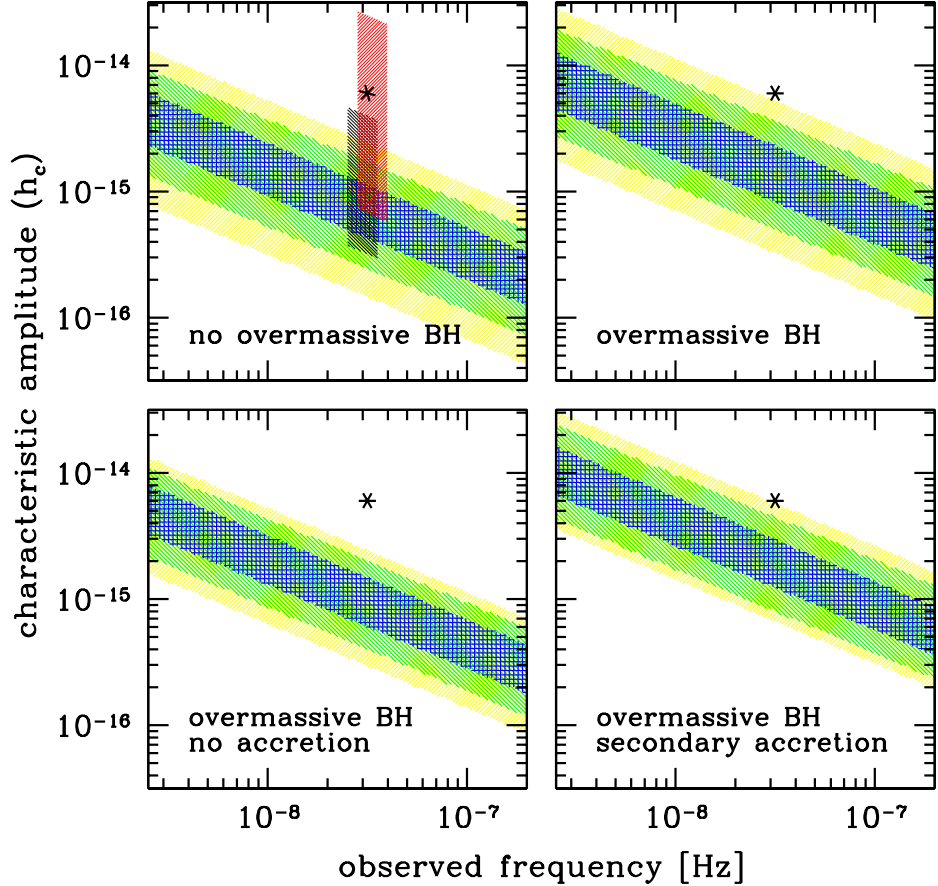}
      \caption{Characteristic amplitude of the GW signal. Shaded areas represent the  $68\%$, $95\%$ and $99.7\%$ confidence levels given by our models. In each panel, the black asterisk marks the best current limit from \protect\cite{vanh11}. Shaded areas in the upper left panel refer to the $95\%$ confidence level given by \protect\cite{mcwilliams12} (red) and the uncertainty range estimated by \protect\cite{sesana08}. See main text for discussion of the individual panels.}
         \label{figbkg}
   \end{figure}

The first, obvious payoff of a PTA detection is the {\it direct} confirmation of the existence of a vast population of sub-pc (to be precise, sub-$0.01$pc) SMBH binaries. From this we learn that (i) binaries efficiently pair on pc scales following galaxy mergers and (ii) stellar and/or gas dynamics is effective in removing energy and angular momentum from the binary, overcoming the 'last parsec problem'. It can still be the case that some level of stalling may occur, delaying the efficient energy loss due to GW emission. Nevertheless, a direct PTA detection will confirm that SMBH binaries eventually coalesce within an Hubble time. The exact level of the signal then depends on the combined ingredients outlined in the previous section, and obviously those cannot be properly constrained all together by measuring one single number. Many degeneracies persist, as we discuss now by referring to the example shown in figure \ref{figbkg}. Here we plot $68\%$, $95\%$ and $99.7\%$ confidence level of the expected characteristic strain, extracted from a large compilation of models featuring different prescriptions for the SMBH binary population, as described in \cite{sesana13}. All models are consistent with several observations of the galaxy mass function, pair fraction, SMBH-host relation etc, as measured at $z<1$, but uncertainties in all such measurements are reflected into a large range of predicted signals. The difference between the top-left and the top-right panel is given by the recent upgrades in the SMBH mass-host relation \cite{mcconnell13,grahamscott13} to include the overmassive black holes measured in BCGs \cite{fabian12}. The range of expected signal is boosted by a factor of two, with the $99.7\%$ confidence level skimming the best limit imposed by current PTA observations \cite{vanh11,demorest13}, which is $A\approx 6\times 10^{-15}$. In the lower panels, we consider two subset of the models featuring these upgraded relations: (i) those in which accretion does not occur prior to binary coalescence, and (ii) those in which accretion precedes the formation of the binary, and is more prominent on the secondary SMBH \cite{callegari11}. In the latter case, binaries observed by PTA are way more massive, and with a larger mass ratio, implying a much larger (by almost a factor of three) signal. If, for instance, a signal with amplitude $A\sim 3\times 10^{-15}$ at a frequency of $1$yr$^{-1}$ is detected, this would support a picture in which SMBHs accrete copious amount of gas {\it before} forming a binary, since SMBHs that do not do so, thus correlating with the merger progenitors, are unable to produce such a strong GW background. However, the more likely $A\sim \times 10^{-15}$ region, can be the result of several combinations of the parameters defining the merging SMBH binary population, and detailed information about each single ingredients will be hard to disentangle.

\section{Spectral shape: environment coupling and eccentricity evolution}
SMBH binaries evolve in a complex, dense astrophysical environment. Forming after galaxy mergers, they sit at the center of the stellar bulge of the remnant, and they are possibly surrounded by massive gas inflows triggered by dynamical instabilities related to the strong variations of the gravitational potential during the merger episode. Accordingly, two major routes for the SMBH binary dynamical evolution have been explored in the literature: (i) gas driven binaries, and (ii) stellar driven binaries. A detailed description of both scenarios is beyond the scope of this contribution; here we consider simple evolutionary routes and assess their impact on the GW signal. 

\subsection{Gas and star driven binaries}
Let restrict ourselves to circular binaries first. Whatever is the driving dynamical mechanism, the emitted GW is always given by equation (\ref{hrms}). What is different is the time spent by the binary at a given frequency, enclosed in the $dt_r/d{\rm ln}f_r$ term. In the GW driven case this is simply
\begin{equation}
\frac{dt_r}{d{\rm ln}f_r} = \frac{5}{64\pi^{8/3}} {\cal M}^{-5/3}f_r^{-8/3}.
\label{e:fdot}
\end{equation}
Combining equations (\ref{e:fdot}) and (\ref{hrms}) yields a contribution to $h_c\propto M_1^{5/6}q^{1/2}f^{-2/3}$. Therefore, integrating over the coalescence rate, the standard $f^{-2/3}$ power law follows. 

A background of stars scattering off the binary, drives its semimajor axis evolution according to the equation \cite{quinlan96}
\begin{equation}
\frac{da}{dt} = \frac{a^2G\rho}{\sigma}H,
\label{adotstar}
\end{equation}
where $\rho$ is the density of the background stars, $\sigma$ is the stellar velocity dispersion and $H$ is a numerical coefficient of order 15. A problem with equation (\ref{adotstar}) is that the SMBH binary efficiently ejects stars from the galaxy core, and the subsequent evolution relies on the pace at which they diffuse into the so called {\it binary loss cone}. As shown by \cite{sesana13}, substituting $\rho_i$ at the binary influence radius ($r_i\approx GM/\sigma^2$) in equation (\ref{adotstar}) corresponds to 'full loss cone at the influence radius', which has to be expected in a complex triaxial environment of a merger remnant, as corroborated by recent numerical simulations \cite{khan11,preto11}. If we consider, for simplicity, an isothermal sphere, we substitute $\rho_i$ in equation (\ref{adotstar}), and we assume $M_{BH}\propto\sigma^5$, we get that in the stellar driven case $dt/d{\rm ln}f\propto f^{2/3}M_1^{2/3}$, which yields to a contribution of the single binary to the GW background of the form $h_c\propto M_1^2qf$. 

In the case of circumbinary disks, things are even more subtle, and the detailed evolution of the system depends on the complicated and uncertain dissipative physics of the disk itself. Here we consider the simple case of a coplanar prograde disk, with a central cavity maintained by the torque exerted by the binary onto the disk. No mass is allowed to flow through the cavity and the mass accumulates at its edge. This scenario admits a selfconsistent, non stationary solution that was worked out by \cite{ipp99}. In this case, the binary evolution rate can be approximated as \cite{ipp99,haiman09}
\begin{equation}
\frac{da}{dt} = \frac{2\dot{M}}{\mu}(aa_0)^{1/2}.
\label{adotgas}
\end{equation}
Here, $\dot{M}$ is the mass accretion rate at the outer edge of the disk, $a_0$ is the semimajor axis at which the mass of the unperturbed disk equals the mass of the secondary black hole, and $\mu$ is the reduced mass of the binary. Considering a standard geometrically thin, optically thick disk model \cite{ss73}, one finds $dt/d{\rm ln}f\propto f^{-1/3}M_1^{1/6}$, which yield to a contribution of the single binary  to the GW background of the form $h_c\propto M_1^{7/4}q^{3/2}f^{1/2}$. 

Compared to the GW driven case, $(da/dt)_{GW}\propto a^{-3}$, equations (\ref{adotstar}) and (\ref{adotgas}) have a very different (milder and positive) $a$ dependence. Therefore, equating  equations (\ref{adotstar}) and (\ref{adotgas}) to $(da/dt)_{GW}$ gives the transition frequency between the external environment driven and the GW driven regimes:
\begin{eqnarray}
f_{{\rm star/GW}}\approx 5\times10^{-9}M_8^{-7/10}q^{-3/10}\,{\rm Hz}\nonumber\\
f_{{\rm gas/GW}}\approx 5\times10^{-9}M_8^{-37/49}q^{-69/98}\,{\rm Hz},
\label{decoup}
\end{eqnarray}
where $M_8=M/10^8\msun$. We therefore see that if the signal is dominated by $10^{9}\msun$ SMBH binaries, then the transition frequency is located around $10^{-9}$Hz. 

\subsection{Eccentricity}
The eccentricity evolution of the binary has a major impact on the GW background through the function $g(n,e)$ \cite{pm63}. The net effect of a large eccentricity is to move power from the second to higher harmonics. However, since the energy carried by the wave is proportional to $f^2h^2$, shifting the emission to higher harmonics effectively {\it removes} power at low frequencies \cite{enoki07}, without a significant enhancement (just marginal) of $h$ at higher frequencies. Therefore, generally speaking, highly eccentric binaries pose a threat to PTA GW detection. 

It is well known that GW emission efficiently circularizes binaries, however things can be drastically different in the star and gas dominated stages. If binaries get very eccentric in those phases, they can retain substantial eccentricity even during the GW dominated inspiral relevant to PTA observations, beyond the decoupling frequencies given by equation (\ref{decoup}). The eccentricity evolution in stellar environments has been tackled by several authors by means of full N-body simulations. Despite the limited number of particles ($N<10^6$), resulting in very noisy behavior for the binary eccentricity, clear trend have been tracked. In general, equal mass, circular binaries tend to stay circular or experience a mild eccentricity increase \cite{mms07}, while binaries that form already eccentric, or with $q\ll 1$ (regardless of their initial eccentricity) tend to grow more eccentric \cite{mat07,preto11}, in reasonable agreement with the prediction of scattering experiments \cite{quinlan96,sesana06}. The same trends were reproduced by \cite{sesana10} exploiting a hybrid model that couples three body scattering experiments of bound and unbound stars to an analytical description of the stellar distribution and of the loss cone refilling. The important parameter here seems to be the eccentricity of the binary at the moment of formation{\footnote{The moment of formation, or pairing, is defined here as the transition point between the early dynamical friction driven stage and the late three body ejection driven stage in the binary evolution. This occurs when the mass in stars enclosed in the binary semimajor axis $a$ is of the order of $M_2$, which corresponds to $a\approx 1-10$pc for the $\sim10^9\msun$ systems of interest here. More details can be found in \cite{sesana10}.}}, $e_0$, which is often found to be larger than 0.6 in numerical studies \cite{preto11}. Large $e_0$ implies that systems emitting in the nHz regime can be highly eccentric, causing a significant suppression of the GW signal, as we will see in the next section. In the circumbinary disk scenario, excitation of eccentricity has been seen in several simulations \cite{armitage05,cuadra09}. In particular, the existence of a limiting eccentricity has been studied in \cite{roedig11} through a suite of high resolution smoothed particle hydrodynamics simulations, in the case of massive selfgravitating disks. They find a critical value $e_{\rm crit} \approx 0.6 - 0.8$. The authors presented an analytical model that agrees with their simulations, predicting the limiting eccentricity to be:  $e_{\rm crit}=0.66 \sqrt{{\rm ln}(\delta - 0.65)}+0.19$, where $\delta\approx R_{c}/a$ is the radius of the disk central cavity in units of the binary semimajor axis. Therefore, also in gaseous rich environments, eccentric binaries might be the norm (even though the extreme eccentricities ($e>0.9$) that might be reached in the stellar driven case are unlikely).

\subsection{What can we learn from PTA observations}
   \begin{figure}
   \centering
   \includegraphics[width=5.0in]{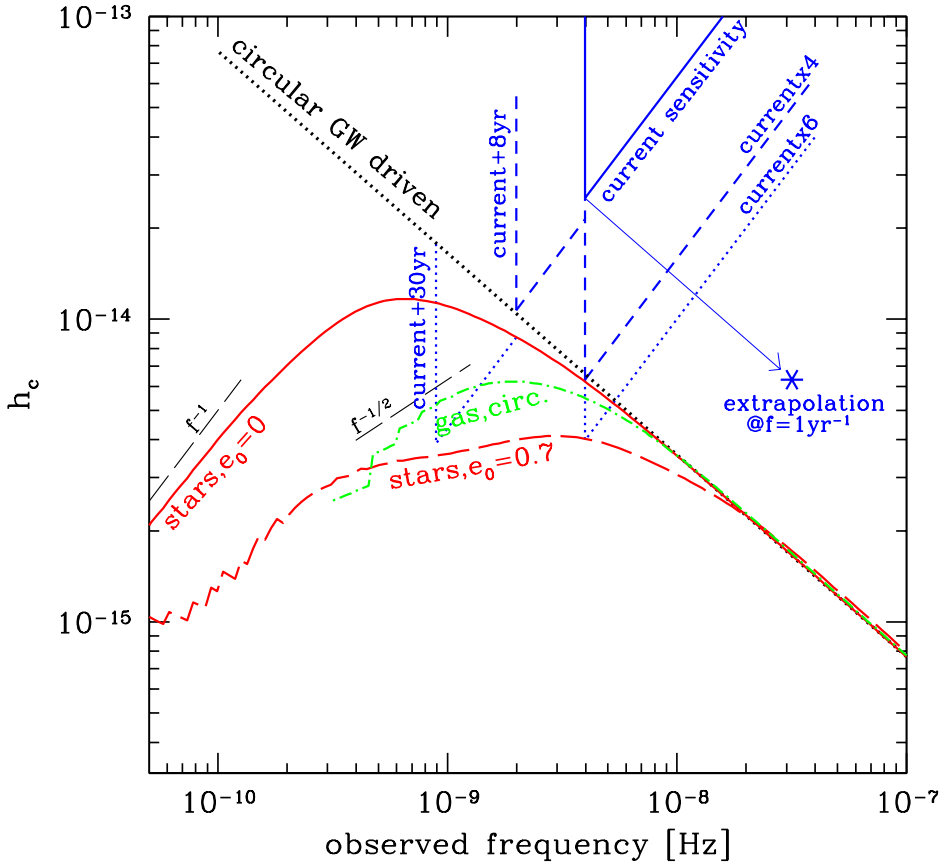}
      \caption{Influence of the binary-environment coupling on the GW signal. The black dotted line is the standard $f^{-2/3}$ spectrum for a population of circular GW driven systems. Red lines are for star driven binaries with eccentricity of 0 (solid) and 0.7 (long-dashed) at pairing; the green dot-dashed line is for circular gas-driven binaries. A sketch of the current PTA sensitivity is given by the solid blue line, which is then extrapolated to the limit at 1 yr$^{-1}$. Also shown in blue are extrapolation of the current sensitivity to include 8 and 30 more years of observations (here we assume no improvement in the timing of the pulsars, the mild improvement in the sensitivity floor is given by the $T^{1/4}$ gain that comes from the longer integration time), as well as the sensitivity given by putative arrays with 4 and 6 time better timing precision. We stress that the sensitivity curves are sketchy and only illustrative, but capture the trends relevant to the discussion in the text.} 
         \label{figobs}
   \end{figure}
The main effect of the binary-environment coupling is to {\it suppress} the low frequency signal \cite{sesana04,enoki07,kocsis11}, as shown in figure \ref{figobs}. The energy of the SMBH binaries is transferred to the environment instead of going into GWs, the binary evolution is faster, and consequently there are less systems emitting at each frequency. The transition frequency (equation (\ref{decoup})) is around $10^{-9}$ Hz (corresponding to 30 yr timescale), but can still have a significant impact on the detection. As depicted in figure \ref{figobs}, PTA limits on the observed background are usually established at very low frequency (given by the timespan of the observations) and then extrapolated at $1$yr$^{-1}$ assuming an $f^{-2/3}$ power law. One might therefore think that, just by keeping observing, PTAs will eventually hit the low frequency background. If the spectral shape changes or, even worse, if there is a turnover frequency, this will not be the case. In figure \ref{figobs} we examine one specific SMBH population model, but we vary the environmental coupling. If we assume roughly the current sensitivity, observing for 8 more years will eventually lead to a detection if SMBH binaries are {\it circular and GW driven}. It might take just a couple of years more if the system are driven by stellar scattering, but it might take some extra 10 years more if all the systems are surrounded by massive circumbinary disks (this clearly depends on the detailed physics of the disk-binary interaction, we just show here a selected case for the sake of the discussion). The situation gets even worse if binaries are eccentric. In particular, we consider the case where SMBH binaries are stellar driven, and all have an eccentricity $e_0=0.7$ at the moment of pairing. The subsequent evolution will result in a population of very eccentric systems generating an almost flat spectrum at $f<3\times10^{-9}$ Hz. With the current timing precision, 30 more years would be needed to detect such signal. It is therefore extremely important for PTAs to constantly improve their intrinsic sensitivity, by reducing timing noise or adding new pulsars, to avoid unpleasant surprises related to SMBH binary dynamics. It is also clear that the determination of the GW background spectral slope carries a lot of information about the dynamics of SMBH binaries. A well defined turnover frequency around $10^{-9}$ Hz will be the distinctive signature that strong coupling with the environment is the norm, whereas a plateau might be indicative of a population of highly eccentric systems. PTA detection will therefore provide important information about the dynamics of individual SMBH binaries, not only about the statistics of their collective population.

\section{Conclusions}
Pulsar timing arrays are achieving sensitivities that might allow the detection of the predicted GW background produced by a cosmological population of SMBH binaries. Beyond the obvious excitement of a direct GW observation, the detection of such signal, together with the determination of its amplitude and spectral slope, will provide an enormous wealth of information about these fascinating astrophysical systems, in particular:
\begin{enumerate}
\item it will give {\it direct unquestionable} evidence of the existence of a large population of sub-parsec SMBH binaries, proving another crucial prediction of the hierarchical model of structure formation;
\item it will demonstrate that the 'final parsec problem' is solved by nature;
\item it will provide important information about the global properties of the SMBH binary population, giving, for example, insights about the relation between SMBH binaries and their hosts;
\item it will inform us about the dynamics of SMBH binaries and their stellar and/or gaseous environment, possibly constraining the efficiency of their mutual interaction;
\item it will tell us if very eccentric SMBH binaries are the norm.  
\end{enumerate}
Identification and sky localization of individual sources (not treated here, see T. Tanaka \& Z. Haiman contribution to this issue), will add further items to this list, making multimessenger studies of SMBH binaries and their hosts possible. Pulsar timing arrays are not mere gravitational wave detectors, but also groundbreaking astrophysical probes that will shed new light on some of the fundamental, yet most elusive objects of our Universe: supermassive black hole binaries.

\section*{Acknowledgments}
A.S. acknowledges the DFG grant SFB/TR 7 Gravitational Wave Astronomy and by DLR (Deutsches Zentrum fur Luft- und Raumfahrt).

\section*{References}
\bibliographystyle{unsrt}
\bibliography{references}

\end{document}